\documentclass[conference]{IEEEtran}

\ifCLASSINFOpdf

\else

\fi

\usepackage{dsfont}
\usepackage{url} 
\usepackage[linesnumbered,ruled,vlined]{algorithm2e}
\usepackage{graphicx}
\usepackage{amsmath} 
\usepackage{amsfonts} 
\usepackage{comment}
\usepackage{hyperref}
\usepackage{algorithmicx}
\usepackage{mdframed} 
\usepackage{xcolor}

\begin{document}

\title{Exploring Green AI for Audio Deepfake Detection}

\author{\IEEEauthorblockN{Subhajit Saha$^{1,3}$, Md Sahidullah$^{1,3}$, Swagatam Das$^{1,2,3}$}
\IEEEauthorblockA{$^1$Institute for Advancing Intelligence, TCG Centres for Research and Education in Science and Technology\\
$^2$Indian Statistical Institute, Kolkata\\
$^3$Academy of Scientific and Innovative Research (AcSIR), Ghaziabad- 201002, India\\
e-mail: subhajit.saha.131@tcgcrest.org, md.sahidullah@tcgcrest.org, swagatam.das@tcgcrest.org}}

\maketitle

\begin{abstract}

The state-of-the-art audio deepfake detectors leveraging deep neural networks exhibit impressive recognition performance. Nonetheless, this advantage is accompanied by a significant carbon footprint. This is mainly due to the use of high-performance computing with accelerators and high training time. Studies show that average deep NLP model produces around 626k lbs of CO\textsubscript{2} which is equivalent to five times of average US car emission at its lifetime. This is certainly a massive threat to the environment. To tackle this challenge, this study presents a novel framework for audio deepfake detection that can be seamlessly trained using standard CPU resources. Our proposed framework utilizes off-the-shelve self-supervised learning (SSL) based models which are pre-trained and available in public repositories. In contrast to existing methods that fine-tune SSL models and employ additional deep neural networks for downstream tasks, we exploit classical machine learning algorithms such as logistic regression and shallow neural networks using the SSL embeddings extracted using the pre-trained model. Our approach shows competitive results compared to the commonly used high-carbon footprint approaches. In experiments with the ASVspoof 2019 LA dataset, we achieve a 0.90\% equal error rate (EER) with less than 1k trainable model parameters. To encourage further research in this direction and support reproducible results, the Python code will be made publicly accessible following acceptance\footnote{\href{https://github.com/sahasubhajit/Speech-Spoofing-}{GitHub link}}.

\end{abstract}


\IEEEpeerreviewmaketitle

\section{Introduction}
\label{Section:Introduction}
\vspace{-0.2cm}
Advancements in deep learning algorithms have revolutionized various fields of machine learning by showcasing impressive performance across different tasks in computer vision (CV), natural language processing (NLP), and speech technology, to name a few~\cite{lecun2015deep}. Arguably, the majority of the improvement stems from the utilization of high-performance computing to process large amounts of image, video, text, and speech data~\cite{prince2023understanding}. The organization of different machine learning challenges also boosted the collaborative efforts within the research community and facilitated systematic benchmarking of algorithms and methodologies. In spite of all these encouraging progress, the carbon footprint and energy consumption are becoming an important concern~\cite{strubell2019energy}. This study highlights environmental concerns of modern deep learning systems and introduces a framework for crafting an eco-friendly, computationally and memory-efficient machine learning system with high performance. It specifically targets its application in detecting audio deepfakes.

Modern deep learning algorithms, especially in the areas of CV, speech, and NLP are predominantly built upon foundation models comprising billions of parameters. Training these algorithms requires massive amounts of training data and often hundreds of GPUs, taking several days to complete. All of these necessitate significant energy consumption and expenditure which are often overlooked as obtaining ``state-of-the-art'' performance is typically the primary focus~\cite{schwartz2020green,Strubell_Ganesh_McCallum_2020}. Leaderboards in machine learning challenges often prioritize showcasing the efficacy of models without duly considering their associated costs or efficiency. The cost for conducting a machine learning experiment to compute the result ($R$), as defined by~\cite{schwartz2020green},

\vspace{-0.5cm}
\begin{equation}
\mathrm{Cost}(R) \propto E \times D \times H,
\label{Eq1}
\end{equation}
\vspace{-0.5cm}

where the training cost $E$ of executing the model on a single example, dataset size $D$, and the number of hyper-parameters $H$. The participants with better computing resources often outperform others and the associated cost as given in Eq.~\ref{Eq1} remained overlooked. Even with alternative options like fine-tuning the pre-trained foundation model, these activities still demands high-performance computing. 

Recently, there's been a push for \textcolor{green}{\textbf{Green AI}}, which is environmentally friendly, as opposed to the traditional \textcolor{red}{\textbf{Red AI}} that relies heavily on massive data and computation~\cite{schwartz2020green,verdecchia2023systematic}. Green AI refers to AI research that achieves new results while considering the computational costs and resources involved. Despite the significant scientific contributions of Red AI, there's been less exploration of the Green AI option, which not only benefits the environment but is also suitable for medium-scale industries and academic institutions with limited budgets for computational resources. This study presents a new Green AI framework for detecting audio deepfakes.

Rapid advancements in neural generation of synthetic speech using \emph{voice conversion} (VC) and \emph{text-to-speech} (TTS) techniques, commonly referred to as audio deepfake technology, have presented a significant threat of identity theft and fraudulent impersonation~\cite{kamble2020advances}. A significant amount of research has been conducted in recent years, largely due to the availability of publicly accessible large datasets from initiatives like the ASVspoof challenge series\footnote{\url{https://www.asvspoof.org/}}, ADD challenges\footnote{\url{http://addchallenge.cn/}}, and others~\footnote{\url{https://deepfake-demo.aisec.fraunhofer.de/in_the_wild}}. Algorithms developed to detect audio deepfakes primarily utilize speech spectrum or other features as the front-end, with the back-end consisting of deep neural networks (DNNs) based on convolutional neural networks (CNNs) or their variants such as ResNet~\cite{liu2023asvspoof}. The key idea is to learn the artifact in synthesized speech and model them using neural networks. A light \emph{convolutional neural network} (LCNN) with \emph{linear-frequency cepstral coefficients} (LFCCs) has been shown state-of-the-art results on several ASVspoof datasets~\cite{wang21fa_interspeech}. Recently, the \emph{audio anti-spoofing using integrated spectro-temporal graph attention networks} (AASIST) approach has shown promising results, utilizing raw waveform as input~\cite{9747766}. The AASIST framework is further improved by integrating with the wav2vec 2.0 based speech foundation model as the front-end, which undergoes joint fine-tuning with the AASIST classifier~\cite{tak22_odyssey}. In recent times, a good amount of research has been conducted in this direction either by exploring other speech foundation models or by combining various levels of information using fusion techniques~\cite{SINHA2024101599,kang2024experimental}. All these efforts help in improving the deepfake detection performance; however, they belong to the Red AI approach and drastically increase the computational overhead and latency.

Our work builds on insights from ongoing Red AI research. It is widely known that speech foundation models are efficient at learning speech patterns to distinguish real from fake speech. We revisit classical machine learning algorithms~\cite{borzi2022synthetic} with those learned representations. We advocate for using the publicly available, widely used speech foundation model for feature extraction without further fine-tuning it~\cite{yang21c_interspeech}. The proposed approach falls under the category of Green AI as it does not necessarily require high-performance computing for training and inference of the deepfake detector. Moreover, we explore layer-wise representations and the potential use of earlier layers of the speech foundation model, which significantly reduces parameters and computational time. To the best of our knowledge, this is the first study to investigate the use of classical machine learning techniques alongside speech foundation models for audio deepfake detection.

The main contributions of this work are summarized as follows:
\begin{itemize}
    \item We employ speech foundation models for speech feature extraction without further fine-tuning.
    \item In stead of using deep models which is widely studied, we thoroughly evaluate the effectiveness of up to six different classical ML algorithms.
    \item Our comprehensive experiments on the LA subset of the ASVspoof 2019 dataset demonstrate that state-of-the-art results can be achieved by considering the initial layer representations from the speech foundation model.
    \item We achieve competitive results with a very few model parameters, and both training and inference computations can be efficiently performed with just single CPU.
\end{itemize}

\vspace{-0.1cm}
\section{Overview of the Proposed Methodology}
\begin{figure}[htp]
    \centering
    \includegraphics[width=8cm]{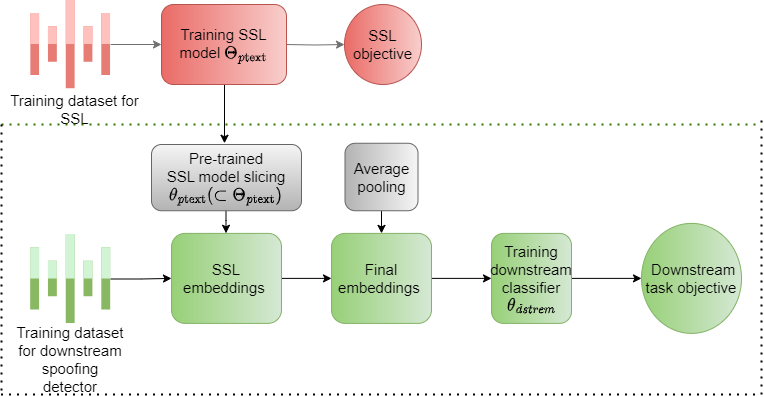}
    \vspace{-0.3cm}
    \caption{Schematic diagram of the proposed framework illustrating SSL training (in \textcolor{red}{red}) and downstream training (in \textcolor{green}{green}).}
    \label{fig:galaxy}
    \vspace{-0.3cm}
\end{figure}

\vspace{-0.1cm}
In this study, we aim to develop a low-carbon footprint methodology for speech anti-spoofing, leveraging the synergies between self-supervised learning (SSL)~\cite{9893562} and traditional supervised machine learning algorithms. While state-of-the-art deep speech foundation models have shown effectiveness in representing raw audio for various tasks, integrating them into downstream tasks requires additional deep learning architectures, and fine-tuning the SSL model for task-specific needs adds complexity by increasing trainable parameters. Our proposed methodology exploits learned representations from various layers of the SSL model, recognizing that each layer contains distinct abstractions of speech. We also propose using aggregated learned representations with a low-cost classification algorithm commonly employed for pattern recognition tasks. As illustrated in Fig.~\ref{fig:galaxy}, our proposed framework consists of two key components: (i)~a \emph{deep feature extractor} based on publicly available pre-trained SSL model, and (ii)~a \emph{downstream classifier} requiring fewer training parameters compared to commonly used DNNs.

Let us denote the two key models, the model trained for pretext task and the downstream task, in the proposed framework as $\Theta_{\mathrm{ptext}}$ and $\theta_{\mathrm{dstream}}$, respectively. Additionally, consider that those two models are trained with $\mathcal{D}_{\mathrm{ptext}}$ (large unlabelled data) and $\mathcal{D}_{\mathrm{dstreamt}}$ (limited unlabelled data), respectively. Usually, the duration of $\mathcal{D}_{\mathrm{ptext}}$ in hours significantly higher than that of $\mathcal{D}_{\mathrm{dstreamt}}$, aligning with the relative differences in parameter counts between $\Theta_{\mathrm{ptext}}$ and $\theta_{\mathrm{dstream}}$. Now we intended to use a \emph{slice} of the pre-trained SSL model which can be denoted by $\theta_{\mathrm{ptext}}$. Therefore, if we only consider first a few layers of the pre-text SSL model, the number of parameters of the $\theta_{\mathrm{ptext}}$ can be substantially lower than the number of parameters in the entire $\Theta_{\mathrm{ptext}}$. This aspect not only reduces the number of parameters but also makes it possible to utilize conventional CPUs for faster inference. The algorithm of the proposed framework can now be summarized in\textbf{ Algorithm~\ref{alg:pretrain_downstream}}.

\vspace{-0.3cm}
\begin{algorithm}
\caption{Summary of the proposed algorithm.}
\label{alg:pretrain_downstream}
\footnotesize 
\begin{algorithmic}[1]

\State\textbf{Pre-training:}
\State Train the model on $\mathcal{D}_{\text{ptext}}$ and estimate $\Theta_{\text{ptext}}$.
\State \textbf{End Pre-training.}
\vspace{0.2cm}
\begin{mdframed}[backgroundcolor=green!10, leftmargin=-0.5em, rightmargin=1em] 
\State \textbf{Downstream feature extraction and modeling:}
\State Compute the features using $\theta_{\text{ptext}} (\subset \Theta_{\text{ptext}})$.
\State Average pooling over frames.
\State Train on $\mathcal{D}_{\text{dstreamt}}$ to get $\theta_{\text{ptext}}$.
\State \textbf{End Downstream Classification.}
\vspace{0.2cm}

\State \textbf{Evaluation:}
\State Compute the features using $\theta_{\text{ptext}}$.
\State Average pooling over frames.
\State Predict the decision using $\mathcal{\theta}_{\text{dstreamt}}$.
\State \textbf{End Evaluation.}
\end{mdframed} 
\end{algorithmic}
\end{algorithm}
\vspace{-0.3cm}

The remaining part of the section contains the description of the adopted SSL and downstream model.

\vspace{-0.1cm}
\subsection{Description of the SSL model}
\vspace{-0.1cm}

In this study, we have considered the wav2vec 2.0 model, which is recognized as one of the most commonly used SSL frameworks for speech tasks~\cite{NEURIPS2020_92d1e1eb}. This consists of a feature encoder, followed by several transformer layers and a quantization module. The feature encoder takes raw waveform as input and consists of a temporal convolution followed by layer normalization and a GELU activation function. The features are processed at the contextual level with several transformer modules. A convolutional layer is used for relative positional embedding. The quantization module discretizes the feature encoder output using product quantization.

We have considered the wav2vec 2.0 \texttt{BASE} model\footnote{\url{https://github.com/facebookresearch/fairseq/tree/main/examples/wav2vec}} with Hugging Face interface\footnote{\url{https://huggingface.co/}} for implementation. The model consists of 12 transformer layers following a CNN with positional encoding layer, which is lightweight compared to the wav2vec 2.0 \texttt{LARGE} model, and requires several days to train even with 128 V100 GPUs. It is also important to note that for processing a speech signal of approximately $3.5$~s (the average size of the training data used in this study), wav2vec 2.0 \texttt{BASE} requires a \emph{multiply-accumulate count} (MAC) of $23.04$~GMACs, while wav2vec 2.0 \texttt{LARGE} requires $60.22$~GMACs. We have extracted the embeddings from each of the transformer layers as well as the feature encoder layer. The dimension of extracted embeddings is $768 \times N$, where 768 is due to the model size of the wav2vec 2.0 \texttt{BASE} model and $N$ is a variable that depends on the number of frames. We compute frame-level average embeddings, i.e., 768-dimensional features, and use them as input to the downstream classifier. 

\begin{figure}[htp]
    \centering
    \includegraphics[width=9cm]{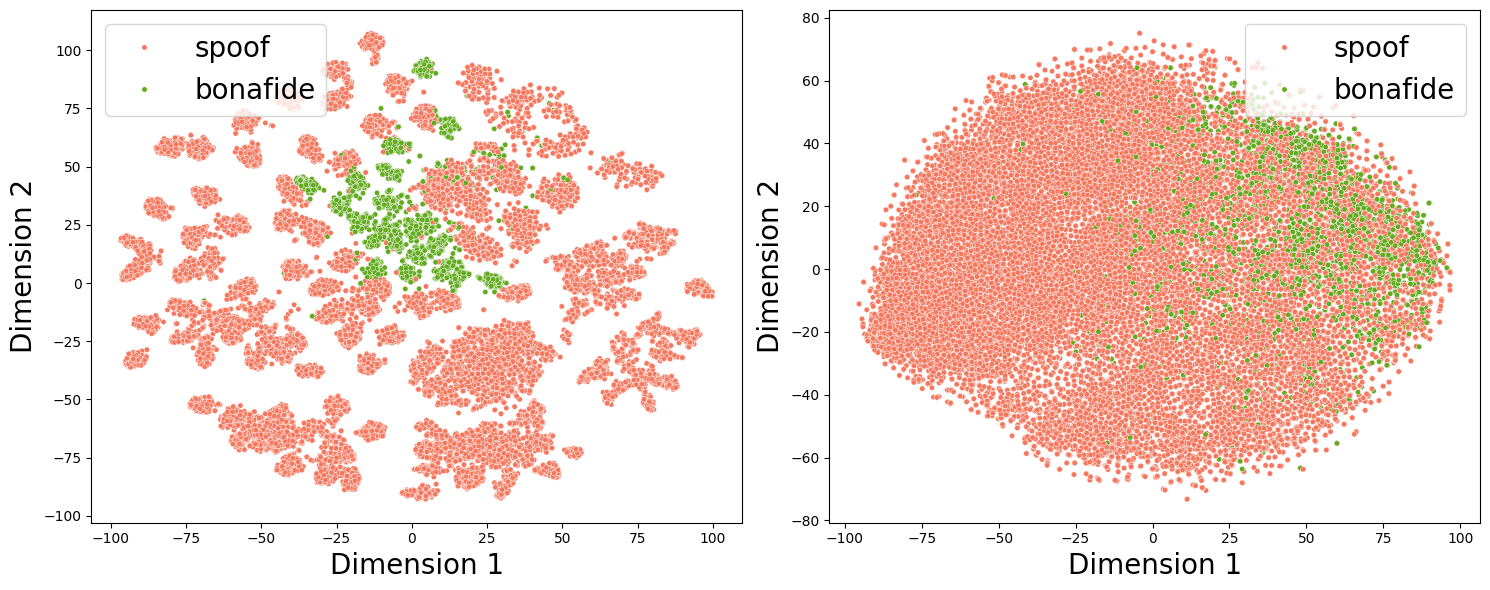}
    \vspace{-0.5cm}
    \caption{t-SNE plots showing embeddings from both the feature encoder output (left) and the 12th layer (last transformer) (right), based on the training set of ASVspoof 2019 LA subset consisting of both bonafide and spoof classes.}
    \label{fig:tsne}
\end{figure}

Figure~\ref{fig:tsne} compares the t-SNE visualization of speech embeddings from two classes computed from various layers of the wav2vec 2.0. Interestingly, the embeddings from the feature encoder layer are visually more distinguishable than the other, which motivates us to explore embeddings from other layers.

\vspace{-0.2cm}
\subsection{Description of downstream classifiers}
\vspace{-0.2cm}
We assess the performance of six traditional classification algorithms listed below. They are capable of being trained efficiently, even on low-end CPUs with limited memory.

\subsubsection{K-nearest neighbors (KNN)}
Our first classification approach employs the k-nearest neighbors algorithm~\cite{duda2006pattern}. This is a non-parametric approach and predicts the class labels of test data by considering the majority class among its k nearest neighbors in the feature space.

\subsubsection{Logistic regression}

Logistic regression is a parametric statistical method used for predicting the probability of an observation belonging to a particular class~\cite{james2013introduction}. It accomplishes this by estimating model parameters that linearly combine input features, resulting in probabilities bounded between 0 and 1. These probabilities are determined through the logistic function. The model parameters are optimized by maximizing the log likelihood function based on the training dataset $\mathcal{D}_{\mathrm{train}}$.

\subsubsection{Support vector machine}

The support vector machine (SVM) finds an optimal decision boundary by maximizing the margin between two classes~\cite{burges1998tutorial}. In this work, SVM employs a kernel trick, using a radial basis function (RBF) kernel, to map the input data into a higher-dimensional feature space where a linear decision boundary can be more easily determined.

\subsubsection{Naive Bayes}
A simple classic ML technique which is used for classification task. It uses Bayes theorem and estimates the posterior probability value of the test sample for each class (also assuming conditional independence between all input features) \cite{Webb2010}. To deal with continuous feature we have used Gaussian Naive Bayes algorithm. 

\subsubsection{Decision tree}

Decision trees are a popular machine learning algorithm known for their simplicity and interpretability~\cite{quinlan1986induction}. They recursively split the data based on some conditions to make predictions or classify instances.

\subsubsection{Multi-layer perceptron}
The multi-layer perceptron (MLP) consists of only one hidden layer~\cite{haykin2009neural}. We have not increased the depth to maintain a lower model complexity.

We have implemented this with the help of the scikit-learn library for Python~\cite{pedregosa2011scikit} utilizing its default parameter settings unless specified otherwise.

\vspace{-0.2cm}
\section{Experimental Setup}
\vspace{-0.2cm}

\vspace{-0.1cm}
\subsection{Dataset}
\vspace{-0.1cm}

We utilized the logical access (LA) subset of the ASVSpoof 2019 dataset for our spoofing detection experiment, as detailed in~\cite{WANG2020101114}. Even though there are several other datasets, this dataset is widely used for spoofing detection research mainly due to its attack diversity and highly controlled spoofing generation and evaluation protocol~\cite{yamagishi2019asvspoof}. The LA subset comprises 19 TTS and VC attacks distributed across training, development, and evaluation subsets. Different statistical and neural methods have been used to generate high quality spoofed data. The training and development sets share the same set of six attacks, whereas the evaluation set comprises 13 unseen attacks. This setup facilitates the evaluation of the generalization ability of the proposed algorithm.

The pre-trained SSL model we employed for this work was trained with 960 hours of audio from the LibriSpeech dataset~\cite{panayotov2015librispeech}. The audio data comprises a read speech corpus in English, similar to the VCTK corpora upon which the ASVspoof 2019 dataset is developed~\cite{WANG2020101114}. Note that although both LibriSpeech and VCTK contain read speech data in English, the latter is recorded in a higher quality studio environment, introducing mismatch in the data conditions.

\vspace{-0.1cm}
\subsection{Hyperparameter settings for downstream classification task}
\vspace{-0.15cm}
  
For downstream classification models, we have conducted a simple brute-force grid search for finding the appropriate hyperparameter setting. We select the values corresponds to the best F1 score on the development set. The Table~\ref{Table1:hyperparameters} describes the key hyperparameter settings of each classification algorithm along with the search spaces. Note that the final chosen hyperparameter is highlighted in boldface for each case.

\begin{table}[h!]
\renewcommand{\arraystretch}{1.3}
\centering
\scriptsize
\caption{Hyperparameters related to grid search for classifiers.}
\vspace{-0.1cm}
\label{Table1:hyperparameters}
\begin{tabular}{|p{1.3cm}| p{6.2cm}|} 
 \hline
 \textbf{Method} & \textbf{Hyperparameter settings} \\ [0.5ex] 
 \hline\hline
  k-NN &  $\mathtt{k=\{3,5,6\}}$\\ 
 \hline

 Logistic regression & $\mathtt{\text{regularization~factor} = \{0.2, 0.1, 10 \}}$ \\
 \hline

 SVM & $\mathtt{\text{regularization~factor} = \{0.2, 0.1, 1\}, kernel = \{rbf\}}$  \\
  \hline
 Decision tree & $\mathtt{\text{split. criteria =  \{gini, entropy\}, max. depth = \{50, 100, 150\}}}$  \\
 \hline
 Naive Bayes & $\mathtt{\text{variance smoothing factor} = 1e^{-9}}$  \\
 \hline
 & $\mathtt{\text{hidden layers = \{(50), (100)\}, activation function = \{relu\}}}$\\
 MLP & $\mathtt{\text{batch size = \{32, 64\}, learning rate = \{constant, invscaling\}}}$ \\
 & $\mathtt{\text{regularization factor = \{0.0001\} }}$ \\ [1ex] 
\hline
\end{tabular}
\vspace{-0.2cm}
\end{table}

\vspace{-0.2cm}
\subsection{Evaluation metrics}
\vspace{-0.2cm}

We have used \emph{equal error rate} (EER) and F1 score metrics for performance evaluation on the test set. This EER is calculated based on a decision threshold where false positive rate and false negative rate are equal. The F1 score is also important to report as the evaluation dataset is imbalanced. We have annotated the bonafide class as positive class. A lower EER and higher F1 score indicate better performance.

\vspace{-0.1cm}
\section{Results}
\vspace{-0.1cm}

\subsection{Comparison of backends}
\vspace{-0.15cm}
 
\begin{figure}[htp]
    \centering
    \includegraphics[width=8cm]{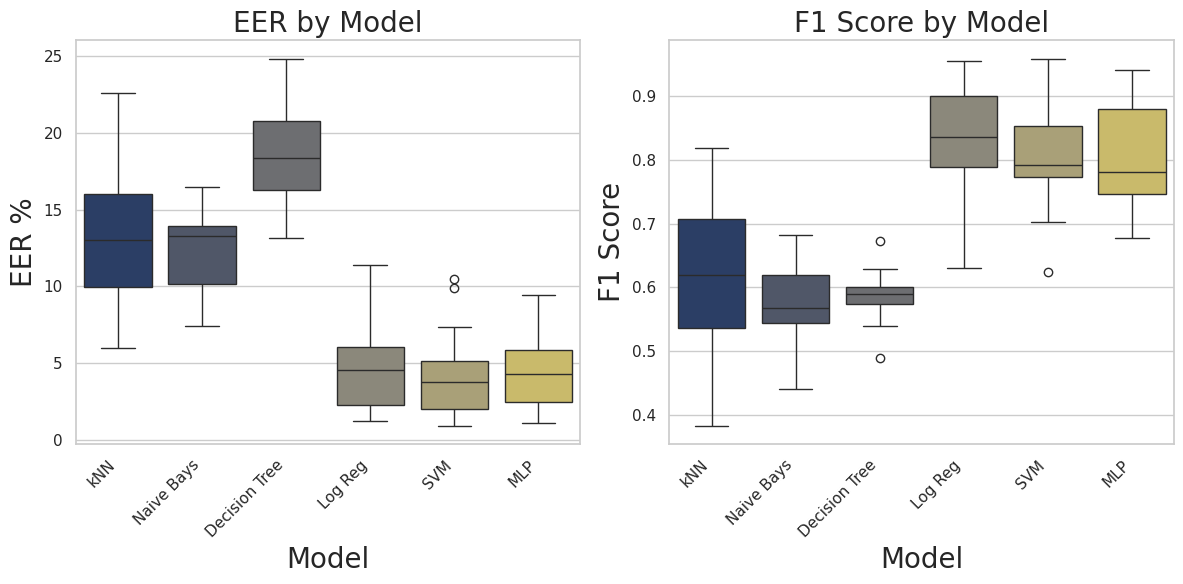}
    \vspace{-0.1cm}
    \caption{Boxplots of EERs (left) and F1 scores (right) for the six methods. Each boxplot summarizes the EERs computed across 13 different layers.}
    \vspace{-0.1cm}
    \label{fig:model}
\end{figure}

In our initial experiment, we compare all six downstream models on the LA subset of ASVspoof 2019. The results are depicted in Fig.~\ref{fig:model}, where we report performance by separately assessing spoofing detection across all 13 layers (the output of one feature encoder layer plus 12 transformer layers). The figure indicates that logistic regression, SVM, and MLP generally outperform the others. We have achieved average EER of 4.25 \% using SVM for downstream task modeling, slightly better than MLP (4.37 \%) and logistic regression (4.87 \%). The best performance with SVM, as separately shown in Table~\ref{tab:compare}, is competitive to the state-of-the-art AASIST model which gives EER of 0.83\% on the same test condition~\cite{9747766}.

Note that these reports are based on the optimal hyperparameter settings for each model and embedding, with each training instance being independent and not utilizing further fusion of models or ensemble techniques.

\vspace{-0.1cm}
\subsection{Effect of choice of layers}
\vspace{-0.1cm}

\begin{figure}[htp]
    \centering
    \includegraphics[width=8cm]{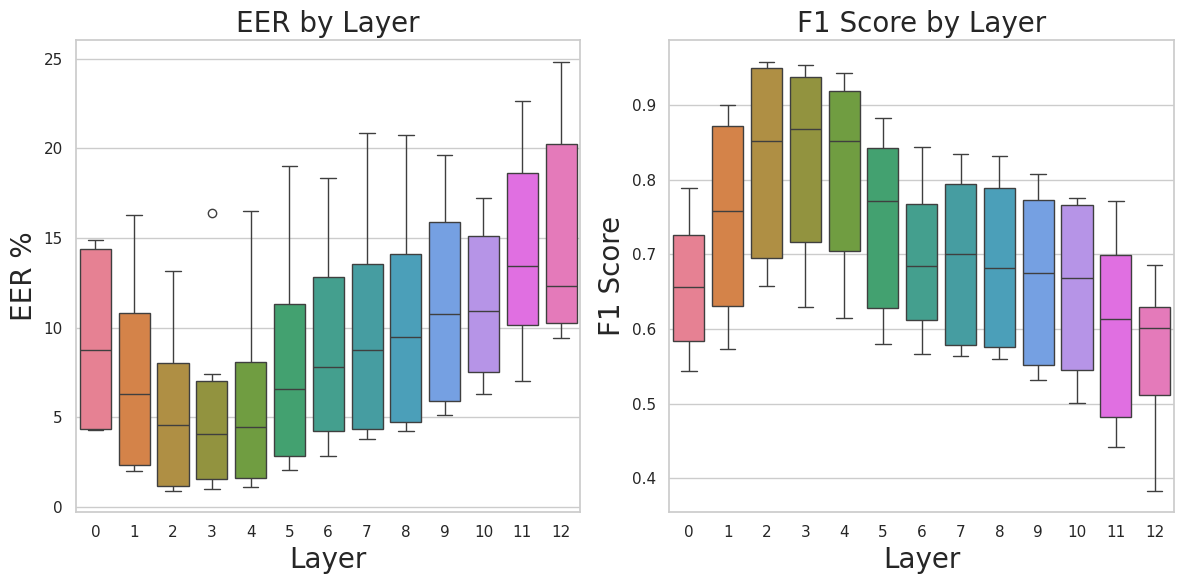}
    \vspace{-0.3cm}
    \caption{Boxplots of EERs and F1 scores for each layer. Each boxplot summarizes the EERs computed across six downsteam classifiers.}
    \vspace{-0.3cm}
    \label{fig:layer}
\end{figure}

In the next experiment, we compare the spoofing detection performance using embeddings from each layer separately. In Fig.~\ref{fig:layer}, we compare the performance by showing a box-plot of results from all the six classifiers. The results indicate that the second transformer layer gives lowest EER and highest F1 score on average. Figures~\ref{fig:model} and~\ref{fig:layer} suggest that embedding at the third layer, combined with either logistic regression, SVM, or MLP, yields better results. This finding contradicts the common practice where the last layer is usually used to extract the speech embeddings from wav2vec 2.0 model for further processing with deep model such as AASIST~\cite{tak22_odyssey}.

\vspace{-0.2cm}
\subsection{Best configuration}
\vspace{-0.15cm}

Based on the observations made so far, we select the the second transformer layer as the front-end, along with an SVM-backend, considering this as the best system. Table~\ref{tab:compare} compares the proposed methods (i.e., wav2vec 2.0 with MLP, Log. reg., and SVM) with recent state-of-the-art techniques. Our proposed best system (wav2vec 2.0) achieves a $0.90\%$ EER and F1 score of $0.95$ with fewer than 1K trainable model parameters. Our method has additional parameters from the pre-trained SSL model, but these are not trainable.

\begin{table}[h]
\renewcommand{\arraystretch}{1.2}
\begin{center}
\scriptsize
\caption{Summary of the different audio deepfake detection methods. \#params indicates the no of trainable parameters only which are associated with energy consumption during training.}
\vspace{-0.2cm}
\label{tab:compare}
\begin{tabular}{|p{3.2cm}|c |c|c|c|} 
 \hline
 \textbf{System} & \textbf{\#params} & \textbf{GPU} & \textbf{Front-end}  & \textbf{EER}\\ [0.5ex] 
 \hline\hline

  RawGAT-ST \cite{inproceedings2}& 437K& Yes & Raw audio & 1.06\\ [0.01ex] 
  \hline

 SENet \cite{zhang21da_interspeech}& 1,100K& Yes & FFT& 1.14\\[0.01ex] 
 \hline

 Res-TSSDNet \cite{unknown}& 350K& Yes & Raw audio& 1.64\\ [0.01ex] 
 \hline

 Raw PC-DARTS \cite{rawinproceedings}& 24,480K & Yes & Raw audio&  1.77\\ [0.01ex] 
\hline

LCNN-LSTM-sum \cite{Wang2021ACS}& 276K& Yes & LFCC& 1.92\\ [0.01ex] 
  \hline

 AASIST-L \cite{9747766}& 85K& Yes & Raw audio & 0.99\\ [0.01ex] 
 \hline

 AASIST \cite{9747766}& 297K& Yes & Raw audio & 0.83\\ [0.01ex] 
\hline
wav2vec 2.0-light-DARTS \cite{wang2022fully}& NA & Yes & Raw audio &  1.09\\ [0.01ex] 
\hline
\hline

\textbf{Wave2Vec 2.0 - MLP~(ours)} 
 & \textbf{77K}& \textbf{No} & \textbf{Raw audio} & 1.11\\ [0.01ex] 
 \hline  
 \textbf{Wave2Vec 2.0 - Log~(ours)}
 & \textbf{769}& \textbf{No} & \textbf{Raw audio} & 1.25\\ [0.01ex] 
 \hline
\textbf{Wave2Vec 2.0 - SVM~(ours)} 
 & \textbf{808}& \textbf{No} & \textbf{Raw audio} & 0.90\\ [0.01ex] 
 \hline 
\end{tabular}
\vspace{-0.2cm}
\end{center}
\end{table}

\vspace{-0.2cm}
\subsection{Towards Green AI solution}
\vspace{-0.15cm}

We achieve optimal spoofing detection performance by utilizing embeddings from the output of the second transformer layer. Consequently, we confine our use of model parameters to this layer and those preceding it. This drastically reduces computation (MAC count by $\approx \text{12 GMAC } (52 \%)$ on average size training data. Moreover, we need to store and utilize $\approx 19 \text{M} (80\% \text{ less})$ parameters instead of $\approx 95 M$ in the original wav2vec 2.0 \texttt{BASE} model. This significantly reduces overall training cost (i.e., $E$ in Eq. \ref{Eq1}) by minimizing cost at downstream feature extraction step (from algorithm \ref{alg:pretrain_downstream}).

We conduct all experiments using both Intel(R) Core(TM) i7-10700 CPU @ 2.90GHz  instead of any GPU-powered computer. Some studies have reported even lower EER; however, they eventually fall under the category of Red AI, as achieving better performance requires larger models~\cite{kang2024experimental}, model fine-tuning and/or joint training~\cite{wang22_odyssey}, and fusion~\cite{SINHA2024101599}, which also necessitate GPUs.

\vspace{-0.2cm}
\section{Conclusions}
\vspace{-0.2cm}
In this study, we advocate for the optimal use of pre-trained speech foundation models for spoofing detection within a CPU-only architecture, aligning with environmentally-friendly as well as cost-effective AI initiatives. Our initial findings with the wav2vec 2.0 model and the ASVspoof 2019 LA dataset demonstrate the potential of classical ML algorithms. By considering speech embedding from an earlier intermediate layer, we have achieved the best spoofing detection results with an SVM classifier using an RBF kernel, which are competitive with the performance obtained using the state-of-the-art graph neural network-based AASIST model. The key advantage of the proposed framework is that developers can utilize publicly available models without the need for GPU resources for training or fine-tuning. The current study can be extended by exploring other large publicly available speech foundation models as extensively evaluated in SUPERB bench-marking for other tasks. Exploring the generalization ability of the proposed approach on different datasets could be another interesting future direction.

\bibliographystyle{ieeetr}
\bibliography{references}

\end{document}